\documentclass[ApJ, twocolumn]{aastex631}

\usepackage{graphicx}
\usepackage{amsmath}
\usepackage{color}
\usepackage[normalem]{ulem}

\newcommand{\SPA}{School of Physics and Astronomy, Monash University, Clayton VIC 3800, Australia}
\newcommand{\OzGravMonash}{OzGrav: The ARC Centre of Excellence for Gravitational Wave Discovery, Clayton VIC 3800, Australia}

\shorttitle{No evidence for a dip in the BBH mass spectrum}
\shortauthors{Adamcewicz et al.}

\begin{document}

\title{No evidence for a dip in the binary black hole mass spectrum}

\author{Christian Adamcewicz}
\email{christian.adamcewicz@monash.edu}
\affiliation{\SPA}
\affiliation{\OzGravMonash}

\author{Paul D. Lasky}
\affiliation{\SPA}
\affiliation{\OzGravMonash}

\author{Eric Thrane}
\affiliation{\SPA}
\affiliation{\OzGravMonash}

\author{Ilya Mandel}
\affiliation{\SPA}
\affiliation{\OzGravMonash}

\begin{abstract}
Stellar models indicate that the core compactness of a star, which is a common proxy for its explodability in a supernova, does not increase monotonically with the star's mass.
Rather, the core compactness dips sharply over a range of carbon-oxygen core masses; this range may be somewhat sensitive to the star's metallicity and evolutionary history.
Stars in this compactness dip are expected to experience supernovae leaving behind neutron stars, whereas stars on either side of this range are expected to form black holes.
This results in a hypothetical mass range in which black holes should seldom form.
Quantitatively, when applied to binary stripped stars, these models predict a dearth of binary black holes with \textit{component masses} $\approx 10 M_\odot - 15 M_\odot$.
The population of gravitational-wave signals indicates potential evidence for a dip in the distribution of \textit{chirp masses} of merging binary black holes near $\approx 10 M_\odot - 12 M_\odot$.
This feature could be linked to the hypothetical component mass gap described above, but this interpretation depends on what assumptions are made of the binaries' mass ratios.
Here, we directly probe the distribution of binary black hole component masses to look for evidence of a gap.
We find no evidence for this feature using data from the third gravitational-wave transient catalogue (GWTC-3).
If this gap does exist in nature, we find that it is unlikely to be resolvable by the end of the current (fourth) LIGO-Virgo-KAGRA (LVK) observing run.
\end{abstract}

\keywords{Black holes (162) --- Compact objects (288) --- Gravitational wave astronomy (675) --- Gravitational waves (678)}

\section{Introduction}\label{sec:introduction}
The formation history of binary black hole (BBH) systems is still highly uncertain \citep[][]{Mapelli_2018, Mandel_2022, Spera_2022}.
However, as the LIGO-Virgo-KAGRA collaboration \citep[LVK;][]{LIGO,Virgo,KAGRA} observes an increasing number of merging BBH systems via gravitational waves, emerging trends in the BBH population's masses, spins, and redshifts are beginning to shed light on these systems' origins.

With the release of the third gravitational-wave transient catalogue \citep[GWTC-3;][]{GWTC3}, a number of informative features have been identified in the distribution of BBH masses.
Namely, BBH component masses roughly follow a descending power-law with an over-abundance of black holes at $\approx 30 M_\odot$ \citep[e.g.][]{Talbot_2018, GWTC3_rnp, Edelman_2023, Callister_2023, Toubiana_2023, Farah_2023, Farah_2023b, Tiwari_2024}, and a likely over-abundance at $\approx 10 M_\odot$ \citep[e.g.][]{Tiwari_2022, GWTC3_rnp, Edelman_2023, Callister_2023, Farah_2023, Tiwari_2024}.
The former (higher-mass) peak may be evidence of pair-instability supernovae in high-mass stars \citep{Heger_2002, Woosley_2015, Talbot_2018, Croon_2023}, though this has been contested \citep[see, for example][]{Stevenson_2019}, or evidence of dynamical formation in globular clusters \citep{Antonini_2023}.
Meanwhile, the latter (lower-mass) peak may point to a considerable fraction of the BBH population originating from the stable mass transfer channel \citep{van_son_2022}, or may be a result of a peak in core compactness for black hole progenitors in a particular mass range \citep{Schneider_2023}.
Some studies also show marginal evidence for a third peak in the mass distribution near $\approx 20 M_\odot$ \citep[e.g.][]{GWTC3_rnp, Toubiana_2023, Edelman_2023, Tiwari_2024}, although other work suggests this feature is not statistically significant \citep[see][]{Farah_2023}.
If real, this feature may also be related to the aforementioned core compactness peak \citep{Schneider_2023}.
The aforementioned features in the BBH mass distribution may also evolve with redshift \citep{Karathanasis_2023, Rinaldi_2024}.

As alluded to above, stellar models suggest that stellar core compactness does not increase monotonically with stellar mass.
Rather, they suggest the existence of a dip in core compactness for a particular core mass range which depends on the star's metallicity, as well as its mass transfer history \citep[see][]{Muller_2016, Schneider_2021, Schneider_2023}.
As lower core compactness favours a supernova explosion \citep{OConnorOtt:2011, SukhboldWoosley:2014, Ertl:2016, Kresse:2021, Schneider_2021}, stars in the mass range of this dip are predicted to explode in supernovae, leaving behind a neutron star.
Meanwhile, stars with masses on either side of this compactness dip are predicted to either avoid explosions altogether and collapse into black holes as so-called `failed supernovae', or to form black holes after weaker explosions and partial fallback.
\cite{Schneider_2021, Schneider_2023} suggest that this gives rise to a gap in the resulting black hole mass distribution.
Specifically, \cite{Schneider_2023} predict that the BBH mass spectrum should exhibit a dearth of systems with component masses in the range $\approx 10 M_\odot - 15 M_\odot$ (see their Fig.~4) if BBH progenitors experience mass transfer in the course of binary evolution \citep[see, for example][]{Mapelli_2018, Mandel_2022, Spera_2022}.\footnote{The theorised range and depth of this gap depend on the progenitor metallicity, mass transfer history, and uncertainty regarding the amount of material retained via fallback.}
Furthermore, \cite{Schneider_2023} propose that this gap in component masses gives rise to a related $\approx 10 M_\odot - 12 M_\odot$ gap in chirp masses
\begin{equation}
    \mathcal{M} = \frac{(m_1 m_2)^{3/5}}{(m_1 + m_2)^{1/5}},
\end{equation}
where $m_1$ and $m_2$ are the component masses of the heavier (primary) and lighter (secondary) black holes in the binary.\footnote{
The chirp mass is more precisely measured from gravitational-wave signatures for low and intermediate mass systems than the component masses.
}
They suggest there is evidence for such a feature in the gravitational-wave data by pointing out a lack of individual events with posterior support for chirp masses in the range $\approx 10 M_\odot - 12 M_\odot$ (see their Fig. 5).
A dearth of events in this range is also supported by data-driven population analyses of the BBH chirp mass distribution \citep{Tiwari_2021, Tiwari_2022, Tiwari_2024, GWTC3_rnp}.
Namely, the population predictive distributions resulting from these analyses show support for a gap within 90\% credible intervals.

However, in order to connect the predicted features in the individual black hole component masses to the chirp mass, one must make additional assumptions about the pairing between component masses in a merging binary.
\cite{Schneider_2023} argue that binary evolution favours nearly equal masses for the components of merging BBH systems: lower-mass black holes associated with the first compactness peak will predominantly be found in binaries with other lower-mass black holes, and similarly for binaries consisting of two black holes from progenitors above the gap.
However, unless this assumption is strictly enforced when analysing the observational data, an inferred $\approx 10 M_\odot - 12 M_\odot$ chirp mass gap cannot be considered a reliable proxy for the $\approx 10 M_\odot - 15 M_\odot$ component mass gap in question (we show this with examples in Appendix~\ref{sec:chirp_mass}).
If we instead want to relax this assumption about mass pairings while testing the \cite{Schneider_2023} mass gap hypothesis, we can do so by looking for a gap in the component mass distributions directly while attempting to infer the pairing, or mass ratio, distribution directly from the data.

A number of studies, again using data-driven population inference methods, find potential evidence for a dearth of black holes with primary component masses in the range hypothesised by \cite{Schneider_2023}: $m_1 \approx 10 M_\odot - 15 M_\odot$ \citep{Tiwari_2022, Tiwari_2024, Edelman_2023, Toubiana_2023, GWTC3_rnp}.
However, some studies using different data-driven analysis techniques suggest there is no evidence for this feature \citep{Farah_2023, Callister_2023, GWTC3_rnp}.
It is worth emphasising that the models in the aforementioned analyses are constructed in terms of primary mass $m_1$ and mass ratio $q$, as opposed to being constructed in terms of the two component masses $m_1$ and $m_2$.
As a result, although some of these analyses find evidence for a dip in the distribution of $m_1$, they imply no such feature in the distribution of secondary black hole component masses $m_2$.

\cite{Disberg_2023} also hypothesise that a dip in core compactness for stars in a particular mass range may produce a gap in the BBH component mass distribution.
However, their estimate for the range of this gap is shifted upwards relative to \cite{Schneider_2023}, and occurs at $\approx 14 M_\odot - 22 M_\odot$.

In this work, we aim to probe the BBH component mass distributions directly, in order to search for evidence of a gap in the hypothesised range ($m_i \approx 10 M_\odot - 15 M_\odot$).
In Section~\ref{sec:model} we describe a mass model for the BBH population with a flexible gap.
In Section~\ref{sec:GWTC3}, we fit this model to the BBH population using GWTC-3 data, finding no evidence for the hypothesised mass gap.
In Section~\ref{sec:future}, we generate and analyse several mock catalogues of BBH events with a gap in masses at $m_i = 10 M_\odot - 15 M_\odot$, in order to estimate when such a feature may become measurable.
We conclude with a discussion of our findings in Section~\ref{sec:discussion}.

\section{Mass-gap model}\label{sec:model}
We construct a phenomenological population model for BBH component masses including a gap, following the framework set out in \cite{Farah_2022} \citep[see also][]{Farah_2023b}.
We start with a one-dimensional function that will be used as the foundation for the component mass distributions $\bar{\pi}(m_i|\Lambda)$.
Here, $\Lambda$ defines the set of all hyper-parameters governing the shape of the model that we aim to infer.
We choose the statistically favoured \textsc{Multi-Peak} model from \cite{GWTC3_rnp} to act as the basis of our one-dimensional mass model.
This model has the flexibility to capture the key features of the BBH mass distribution outside of the gap range, as is evident when cross-checking with the flexible data-driven fits to the mass distribution from \cite{GWTC3_rnp}.
It consists of a power-law, two Gaussian peaks, low and high mass cutoffs, and a smoothing function at the low mass cutoff.
We then add a flexible gap to this model using a notch filter with hyper-parameters governing the depth $A$, lower-edge location $\gamma_\mathrm{low}$, and upper-edge location $\gamma_\mathrm{high}$.\footnote{
The notch filter has two additional hyper-parameters $\eta_\mathrm{low}$ and $\eta_\mathrm{high}$ that govern the sharpness of the lower and upper edges respectively.
}
We leave the unwieldy mathematical formalism of this function for Appendix~\ref{sec:pi_m}.
The depth $A$ ranges from $0$ (no dip at all) to $1$ (a completely empty gap).

Using this function as a basis, we can construct a two-dimensional mass model for primary mass $m_1$ and secondary mass $m_2$ as
\begin{align}
    \nonumber
    \pi(m_1,m_2|\Lambda) \propto & \
    \bar{\pi}(m_1|\Lambda)
    \bar{\pi}(m_2|\Lambda) \times \\ & \
    f_p (m_1, m_2|\beta)
    \Theta(m_1 - m_2),
\end{align}
where $\Theta(m_1-m_2)$ is a Heaviside step function enforcing $m_1 \geq m_2$, and
\begin{equation} \label{eq:mass_pairing}
    f_p(m_1,m_2|\beta) =
    \left(\frac{m_2}{m_1}\right)^\beta
\end{equation}
is a pairing function that adds some flexibility to the mass ratio distribution.
Again, we make this decision to be somewhat agnostic as to how BBH masses pair with one another.
In the case that $\beta = 0$, BBH masses pair randomly.
On the other hand, as $\beta$ becomes larger, black holes tend to pair with more similar masses.
The latter case approaches the assumption made by \cite{Schneider_2023}: that black holes tend to pair with similar masses, with pairing across gaps in the mass distribution unlikely.
Note that $\beta$ is a subset of $\Lambda$.

Note that while $\bar{\pi}(m_i|\Lambda)$ is used in the construction of this model, this does not give the exact functional form of the marginal distributions due to the introduction of the pairing and Heaviside step functions:
\begin{equation}
    \pi(m_i|\Lambda) = \int_{m_{\min}}^{m_{\max}} dm_j \pi(m_i,m_j|\Lambda)
    \neq \bar{\pi}(m_i|\Lambda).
\end{equation}
Similarly, the mass hyper-parameters do not have the exact same interpretation relative to the marginal mass distributions $\pi(m_i|\Lambda)$ as they do relative to the formational function $\bar{\pi}(m_i|\Lambda)$.
Regardless, the features in either function are qualitatively similar (both consist of a power-law-like basis, two peaks, and a dip from approximately $\gamma_\mathrm{low}$ to $\gamma_\mathrm{high}$ with depth $A$).\footnote{
That is to say, assuming $\gamma_\mathrm{low} = 10 M_\odot$ and $\gamma_\mathrm{high} = 15 M_\odot$ with sharp roll-offs $\eta_\mathrm{low} = \eta_\mathrm{high} = 50$, and drawing samples from two distributions that are identical apart from one having $A=0$ and the other having $A=0.5$, the latter distribution will result in roughly half as many samples with values of $m_i$ between $10 M_\odot - 15 M_\odot$ when compared to the former.
If we suppose a third case with $A=1$, no samples are drawn in this region.
}
We list the subset of the model hyper-parameters $\Lambda$ governing the mass distribution, along with their priors used during hierarchical inference, in Appendix~\ref{sec:pi_m}.

We simultaneously model the spin distribution using the \textsc{Default} spin model from \cite{GWTC3_rnp} \citep[see also][]{Wysocki_2019, Talbot_2017} and the redshift distribution using the \textsc{Power-Law} redshift model from \cite{Fishbach_2018}.

\section{Looking for a gap in GWTC-3}\label{sec:GWTC3}

Using our mass-gap model, we perform hierarchical Bayesian inference in order to infer the mass-gap hyper-parameters from GWTC-3 data.
Our dataset includes the 69 BBH observations that were considered reliable for inclusion in the LVK's GWTC-3 rates and populations analysis paper \citep[events with a false alarm rate $< 1 \mathrm{yr}^{-1}$;][]{GWTC3_rnp}.
We perform our inference using the \texttt{GWPopulation} package \citep{gwpopulation}.
This is built on top of the inference package \texttt{Bilby} \citep{bilby, bilby2} and utilises the nested sampler \texttt{DYNESTY} \citep{dynesty}.
We account for mass, redshift, and spin-based selection effects using recovered injections \citep[see][]{Messenger_2013, Tiwari_2018, Thrane_2019, Farr_2019, Mandel_2019, Essick_2022}.
The GWTC-3 BBH posterior samples and the injection sets used for this analysis originate from \cite{GWTC3, GWTC3_rnp} and are publicly available as \cite{injections, pe}.

\begin{figure}
    \centering
    \includegraphics[width=\columnwidth]{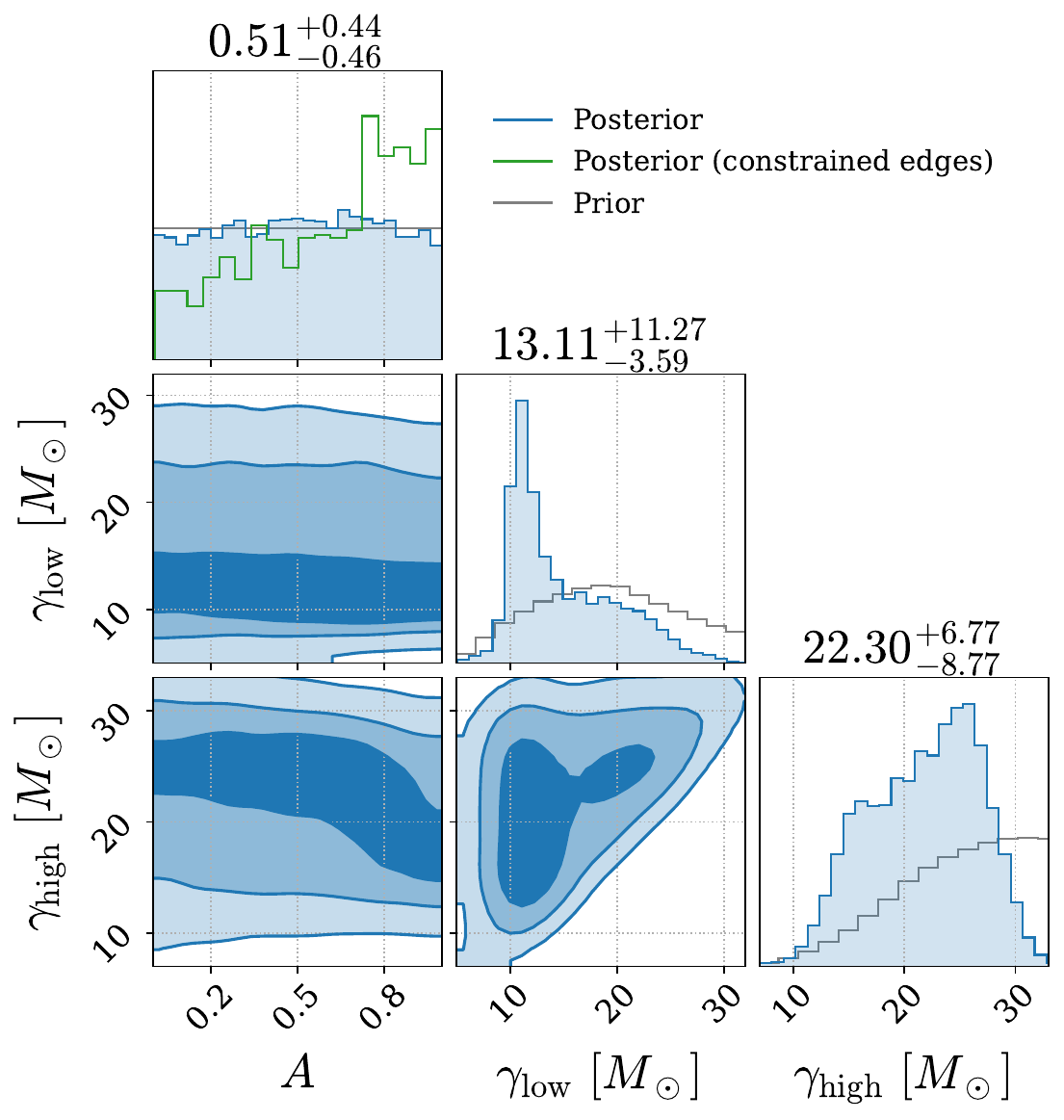}
    \caption{Posterior distributions for gap-related population hyper-parameters inferred using GWTC-3 data (blue).
    The values listed above give the median and 90\% credible intervals on these posteriors.
    The contours on the two-dimensional panels give the 50\%, 90\%, and 99\% credible intervals.
    The green line shows the posterior for the gap depth $A$ when constraining the gap edges to $\gamma_\mathrm{low} = 10^{+1}_{-1} M_\odot$ and $\gamma_\mathrm{low} = 15^{+1}_{-1} M_\odot$.
    The priors are over-plotted in gray.
    Note that there are restrictions that $\gamma_\mathrm{low} > \mu_1$ and $\gamma_\mathrm{high} < \mu_2$.
    These peak locations ($\mu_1$ and $\mu_2$) are well-measured.
    }
    \label{fig:corner}
\end{figure}

We first compare the mass-gap model to an (otherwise identical) model without the flexibility for a gap (gap depth fixed at $A=0$).
For comparing models, we find that the Bayes factor can be misleading due to large regions of zero-probability in the edge-location posteriors.
We instead define a one-dimensional Bayes factor
\begin{equation} \label{eq:B_A}
    \mathcal{B}_A =
    \frac{\int dA \ \mathcal{L}(d|A)}{\mathcal{L}(d|A=0)},
\end{equation}
where $\mathcal{L}(d|A)$ is the population likelihood for the data $d$, given the gap depth $A$ within the gap model.
In practice, this value is computed entirely using the marginal posterior distribution of $A$ in the gap model (e.g., the top left panel in Fig.~\ref{fig:corner}).
We fit the distribution with a Gaussian KDE.
The height of the distribution as $A \rightarrow 0$ is the denominator, and the average height of the distribution integrated over all values of $A > 0$ is the numerator.
This differs from a standard Bayes factor between the no-gap and gap models, as in this case, the Occam's penalty incurred by marginalising over the edge location and sharpness parameters is also applied to the no-gap ($A \rightarrow 0$) scenario.
We find a gap is favoured by a factor of only $\ln \mathcal{B}_A = 0.1$.
Furthermore, comparing the difference in maximum natural log likelihood achieved by a model with a gap, to that achieved by a model without a gap, we find $\Delta \ln \mathcal{L}_{\max} = 1.0$ in favour of the gap model.\footnote{
Note that one can show the expectation value of the difference between the maximum and true log likelihoods scales proportionally with the number of model parameters.
Furthermore, the maximum log likelihood achieved by an analysis is subject to additional uncertainty from the sampling method.
These considerations make $\Delta \ln \mathcal{L}_{\max}$ less reliable as a metric than $\mathcal{B}_A$.}
In the gap model, this maximum natural log likelihood is associated with a gap depth of $A=0.8$.

Note that the gap model (and the gap scenario considered when computing $\mathcal{B}_A$) includes any value of $A > 0$, ranging from very small dips all the way to an empty gap at $A = 1$.
Setting aside the potential for a partially-filled dip, and instead comparing only the cases that there may exist no gap $A = 0$, and that there may exist a completely empty gap $A = 1$, we define a metric
\begin{equation}
    \mathcal{B}_\mathrm{empty} = \frac{\mathcal{L}(d|A=1)}{\mathcal{L}(d|A=0)}.   
\end{equation}
This is computed in the same manner as equation~\ref{eq:B_A}, where the numerator is now proportional to the height of the marginal posterior of $A$ when $A = 1$.
In GWTC-3, we find that the $A = 0$ and $A = 1$ scenarios are indistinguishable, with $\ln \mathcal{B}_\mathrm{empty} = 0.0$.

We plot the posterior distributions for the gap depth, $A$, and location, $\gamma_\mathrm{low}$ and $\gamma_\mathrm{high}$, in Fig.~\ref{fig:corner}.
We find that the posterior on the gap depth $A$ recovers the prior.
The gap edges are restricted to lie between the model's two Gaussian peaks (see Appendix~\ref{sec:pi_m} and Fig.~\ref{fig:pop}), which are relatively well constrained at $\approx 10 M_\odot$ and $\approx 30 M_\odot$.
This appears to be responsible for most of the difference between the posterior and prior for $\gamma_\mathrm{low}$ and $\gamma_\mathrm{high}$.
We find that if the upper-edge of the gap $\gamma_\mathrm{high}$ is restricted to be below $\approx 20 M_\odot$, the data prefers a gap -- albeit with low significance ($A=0$ is still within the 90\% credible interval; see the bottom-left panel of Fig.~\ref{fig:corner}).
Quantitatively, if we slice through the posterior at $\gamma_\mathrm{low} = 10^{+1}_{-1} M_\odot$ and $\gamma_\mathrm{low} = 15^{+1}_{-1} M_\odot$ \citep[asserting the approximate gap location predicted by][]{Schneider_2023}, we find that preference for a gap increases to $\ln \mathcal{B}_A = 0.6$.
Similarly, when the gap location is restricted, we find a value of $\ln \mathcal{B}_\mathrm{empty} = 1.0$ slightly favouring a completely empty gap over no gap at all.
The posterior on the depth with the edge locations constrained is shown in green in Fig.~\ref{fig:corner}.
We plot the posteriors for other hyper-parameters governing the mass distribution in Appendix~\ref{sec:additional_results}.

In Fig.~\ref{fig:pop}, we plot the population predictive distributions for $m_1$ and $m_2$ from our mass models.
There is limited support for a dip in component masses near $\approx 10 M_\odot - 15 M_\odot$ when the gap edges are constrained to this region.
This support disappears when the gap edges are free to lie anywhere between the two peaks.
In the no-gap case, we find that the fraction of primary and secondary mass black holes in the $10 M_\odot - 15 M_\odot$ range is $0.25^{+0.11}_{-0.09}$ and $0.10^{+0.08}_{-0.05}$ respectively.
When a gap is allowed, these fractions drop slightly to $0.24^{+0.11}_{-0.09}$ and $0.09^{+0.08}_{-0.05}$ respectively.
Finally, when the edges of the gap are constrained to the hypothesised region, these fractions fall further to $0.21^{+0.10}_{-0.09}$ and $0.07^{+0.07}_{-0.04}$ respectively.

\begin{figure*}
    \centering
    \includegraphics[width=0.95\textwidth]{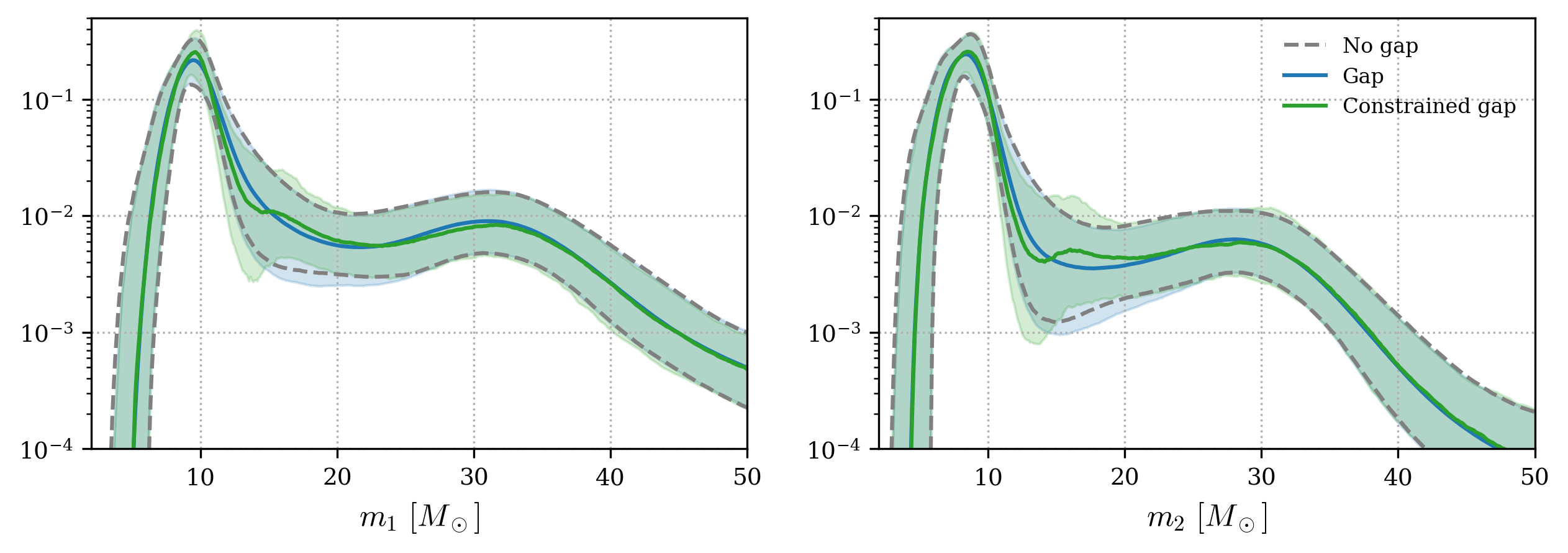}
    \caption{Population predictive distributions for primary mass $m_1$ (left) and secondary mass $m_2$ (right) black holes fit to GWTC-3 data.
    Blue shows the inferred distribution for the standard gap model, while green shows the inferred distribution when the gap's edge locations are constrained to $\gamma_\mathrm{low} = 10^{+1}_{-1} M_\odot$ and $\gamma_\mathrm{low} = 15^{+1}_{-1} M_\odot$.
    The solid lines show the median distributions, while shaded regions give 90\% credible intervals.
    For comparison, the 90\% credible intervals inferred using a model with no gap are shown in gray dashed lines.
    Note the logarithmic vertical scale.}
    \label{fig:pop}
\end{figure*}

\section{When might we see a gap?}\label{sec:future}
Following these inconclusive results, a natural question arises: if there is truly a gap in black hole masses from $\approx 10 M_\odot - 15 M_\odot$, when will it become possible to confidently resolve this feature?

To investigate this, we generate a set of five mock catalogues, each with 250 BBH events.
This is a loose estimate of the number of false alarm rate $< 1 \mathrm{yr}^{-1}$ detections expected to be obtained by the end of the current LVK observing run O4 \citep[see][although, more updated estimates can be made by following the \href{https://gracedb.ligo.org/superevents/public/O4/}{public alert database}]{Petrov_2022, Kiendrebeogo_2023}.\footnote{
We find that catalogues larger than this become exceedingly computationally expensive to analyze.
}
The masses of these mock events are drawn from our mass-gap model with a completely empty gap ($A=1$), with edge locations $\gamma_\mathrm{low}=9.5 M_\odot$ and $\gamma_\mathrm{high}=15.5 M_\odot$ \citep[e.g., expanding slightly on estimates from][]{Schneider_2023}.\footnote{
The gap edges have sharp roll-offs ($\eta_\mathrm{low} = \eta_\mathrm{high} = 50$).
}
Meanwhile, other hyper-parameters governing the shape of the mass distribution and Power-Law redshift distribution are set to match those inferred from GWTC-3 data.
We do not model the spins of the BBH systems in these mock catalogues (see more on this below).
We plot the masses of the events in each mock catalogue in Fig.~\ref{fig:cats_plot} in Appendix~\ref{sec:additional_results}.
We simulate posterior samples for each event using the \texttt{GWMockCat} package \citep{Farah_2023}, along with a matching injection set to estimate search sensitivity.

\texttt{GWMockCat} simulates posteriors for mock events by first drawing mass and redshift parameters from the asserted ``true'' distribution.
The package then assigns an ``observed'' signal-to-noise ratio (SNR) to each draw by first computing an optimal SNR using these drawn parameters and a power spectral density (PSD) representative of typical LVK detector noise.
A random multiplicative factor $\Theta \in [0,1]$ is then applied to model suppression of the SNR due to sky position and orientation \citep[assumed to be isotropic;][]{Finn_1993}, and finally the ``observed'' SNR is assumed to be normally distributed around this ``true'' value.
Events are ``detected'' and added to the catalogue if they are above the assigned threshold (SNR = 8 for single-detector events, as we are assuming here).
Working in terms of source-frame chirp mass $\mathcal{M}$, symmetrical mass ratio $\eta = m_1 m_2 / (m_1 + m_2)^2$ and the multiplicative factor $\Theta$, \texttt{GWMockCat} then assigns ``observed'' parameters for each event that are normally distributed around the true parameters with a standard deviation inversely proportional to the observed SNR.
Finally, \texttt{GWMockCat} generates mock posterior samples for each event by drawing samples (in terms of $\mathcal{M}$, $\eta$, $\Theta$ and the SNR) from a normal distribution centred on the ``observed'' value with the same standard deviation as above.
After performing a coordinate transform, this produces semi-realistic posteriors in component masses $m_i$ and redshift $z$, as verified in Appendices~B~and~C in \cite{Farah_2023}.
Although spin parameters are neglected in this process, the mass and redshift posteriors produced by \texttt{GWMockCat} are similar to real mass and redshift posteriors when marginalising over uncertainty in spins.
The effect of covariance between spin parameters and mass and redshift parameters can generally be considered negligible \citep{Farah_2023}.
For more information on the process that \texttt{GWMockCat} employs, see Appendix~A of \cite{Farah_2023}.

Using these newly generated mock catalogues containing a gap from $\approx 10 M_\odot - 15 M_\odot$, we fit our mass-gap model to probe for evidence of the feature, as was done for GWTC-3 data in Section~\ref{sec:GWTC3}.

\begin{figure}
    \centering
    \includegraphics[width=\columnwidth]{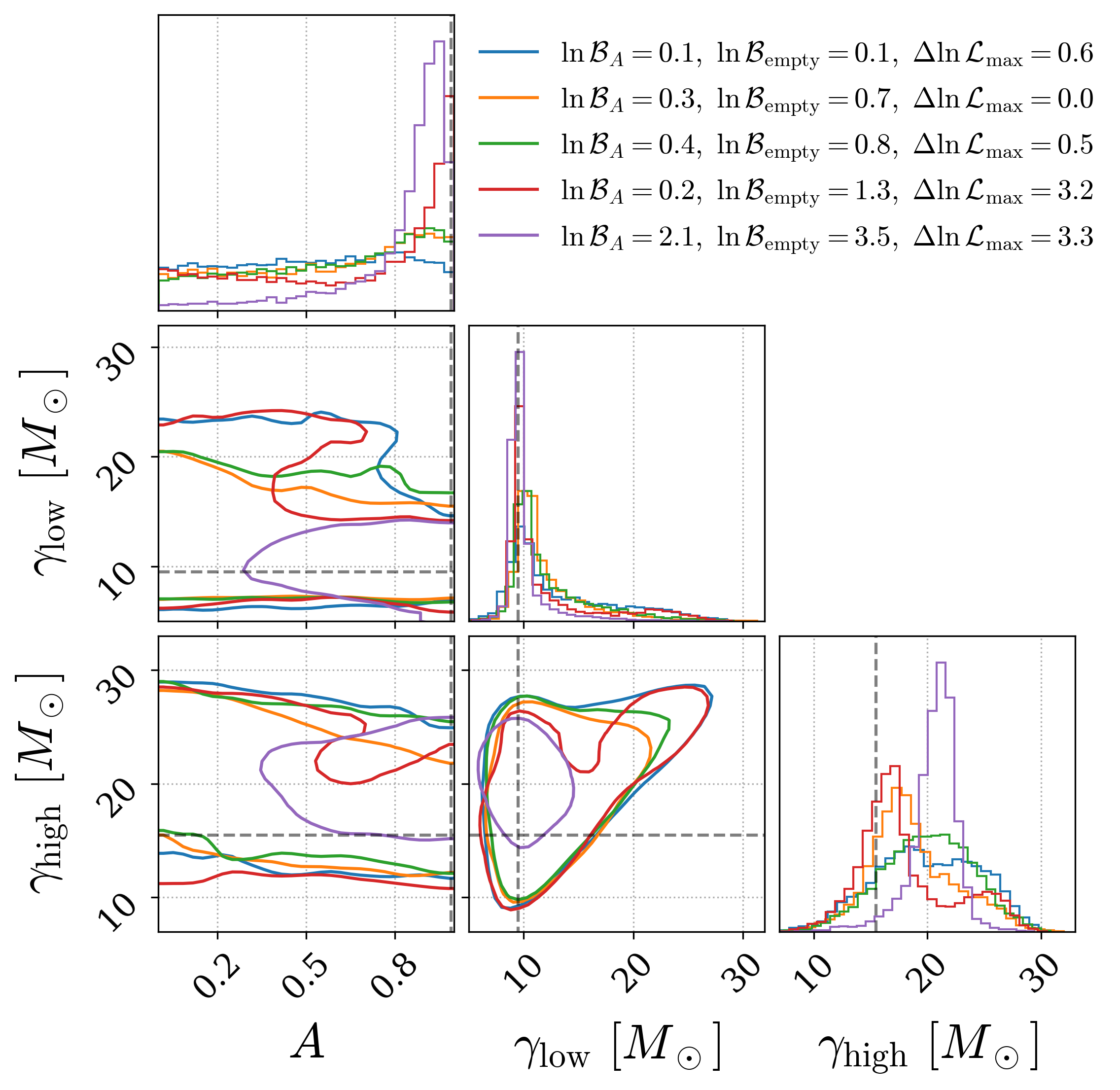}
    \caption{Posterior distributions for gap-related population hyper-parameters inferred from mock catalogues with 250 events each.
    Each colour shows the results for an individual catalogue, while the true values are shown by gray dashed lines.
    The contours on the two-dimensional panels give the 90\% credible intervals.
    The legend gives the natural log one-dimensional Bayes factor, natural log Bayes factor between the empty and no gap hypotheses, and difference in maximum natural log likelihood comparing the gap and no-gap hypotheses for each mock catalogue.}
    \label{fig:mock_corner}
\end{figure}

We plot the inferred gap parameters for the mock catalogues in Fig.~\ref{fig:mock_corner}, also listing the one dimensional Bayes factors $\mathcal{B}_A$, the empty-gap to no-gap Bayes factor $\mathcal{B}_\mathrm{empty}$ and differences in maximum natural log likelihoods $\Delta \ln \mathcal{L}_{\max}$ comparing the gap hypothesis to the no-gap hypothesis for each catalogue.
As expected, $\mathcal{B}_\mathrm{empty} \geq \mathcal{B}_A$ for each catalogue, as $\mathcal{B}_\mathrm{empty}$ does not suffer an Occam's penalty when marginalising over values of $0 < A < 1$.
Regardless, even if real, it appears as if such a gap remains difficult to measure with a catalogue of 250 events.

This difficulty is partly driven by a combination of the uncertainties in individual event posteriors and the relatively high merger rates for binaries with component masses just outside of the gap range, around $\approx 10 M_\odot$.
Individual mass posteriors in this range have characteristic uncertainties of between $\pm 1 M_\odot$ and $\pm 10 M_\odot$ -- similar to the true width of the supposed gap \citep[see, for example, Table IV from][]{GWTC3}.
If $\gtrsim 50\%$ of all merging BBH component masses fall into the peaks surrounding the gap (see Fig.~\ref{fig:full_corner} in Appendix~\ref{sec:additional_results}) and a significant fraction of these are inferred to have masses consistent with gap values due to measurement uncertainty, these can dominate over the $\lesssim 10\%$ of merging BBH systems that would be needed to fill in the gap, making gap detection challenging.
The gap can only be confidently detected if the Poisson scatter on the accidental contribution from surrounding-peak black holes is at least a few times smaller than the number of ``missing'' events excised by the gap.
This would notionally require $\sim 1000$ events when marginalising over an uncertain gap location, though the actual required number is likely greater because this simple estimate does not account for gravitational-wave selection effects that disfavour observations of lower mass BBH systems.

\begin{figure}
    \centering
    \includegraphics[width=0.9\columnwidth]{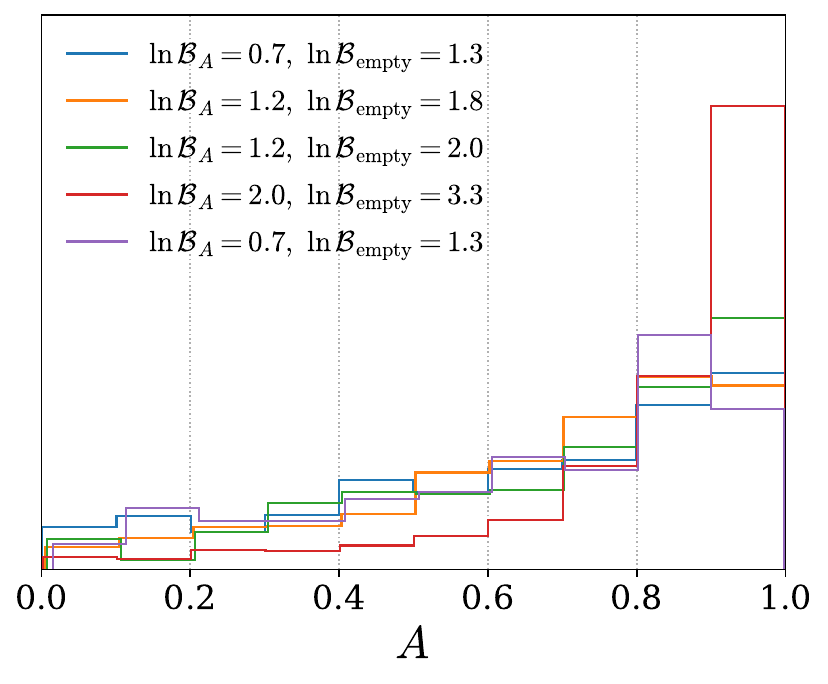}
    \caption{Posterior distributions for the gap depth inferred with each mock catalogue, constraining the gap location to $\approx 10 - 15 M_\odot$.
    Each colour shows the results for an individual catalogue.
    The legend gives the natural log one-dimensional Bayes factor comparing the gap and no-gap hypotheses, as well as the empty-gap and no-gap hypotheses for each mock catalogue.}
    \label{fig:mock_corners_fixed}
\end{figure}

As seen in Fig.~\ref{fig:mock_corner}, the edge locations of the gap in the simulated catalogues tend to be more precisely measured than in the case of GWTC-3.
We also observe variance in the inferred upper-edge $\gamma_\mathrm{high}$ relative to the lower-edge $\gamma_\mathrm{low}$ between the catalogues (although all inferences remain consistent with the injected value).
Again, discerning between the gap and no-gap hypothesis is made more difficult when the location of the gap is not known.
If we constrain the gap edges to lie between $\gamma_\mathrm{low} = 10^{+1}_{-1}$ and $\gamma_\mathrm{low} = 15^{+1}_{-1}$ (as was done in Section~\ref{sec:GWTC3}), we find improved evidence for a gap in most instances.
This is illustrated in Fig.~\ref{fig:mock_corners_fixed}, in which we show the posterior on the gap depth for each catalogue when the gap edges are constrained, and list the corresponding Bayes factors.

\begin{figure}
    \centering
    \includegraphics[width=\columnwidth]{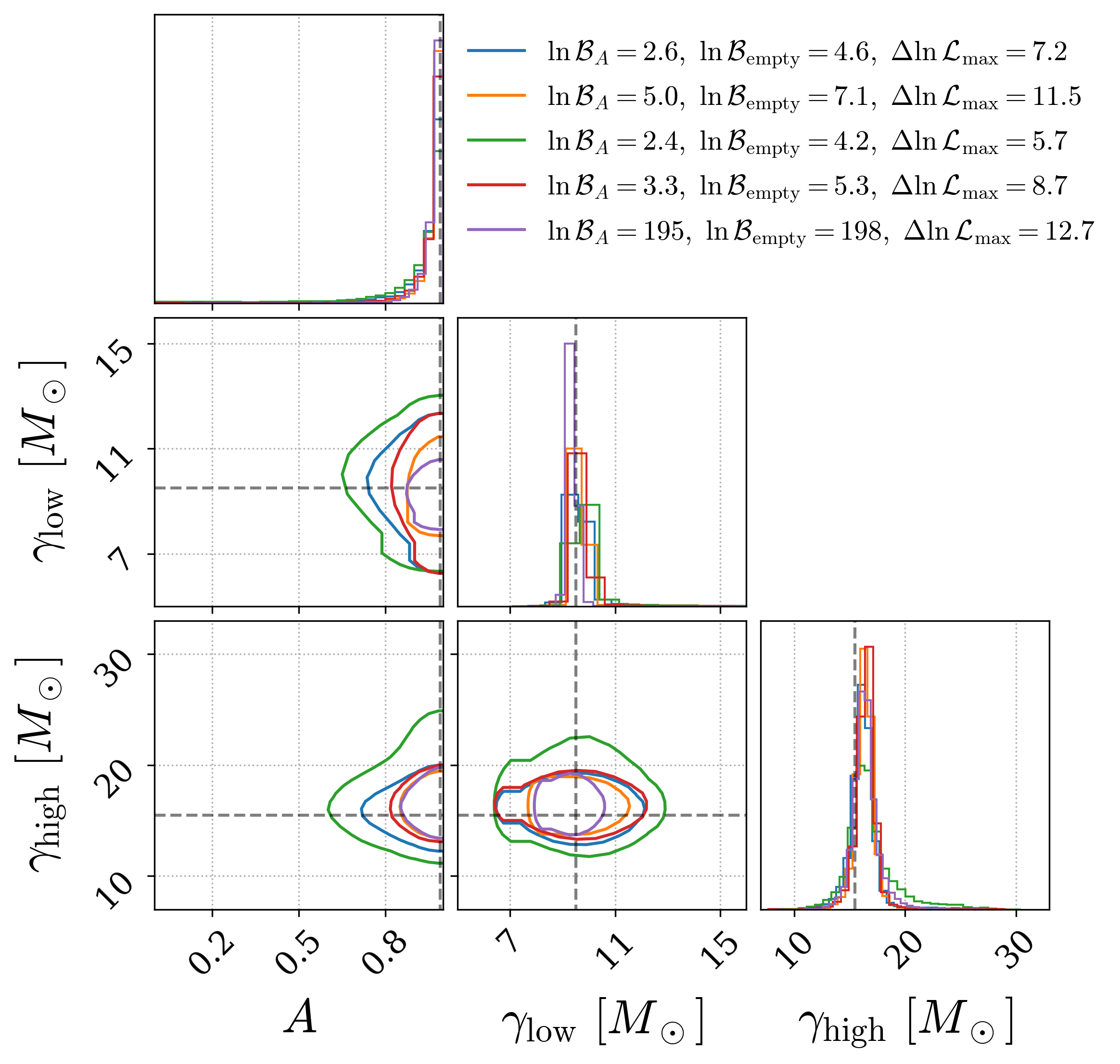}
    \caption{Posterior distributions for gap-related population hyper-parameters inferred from mock catalogues with 250 \textit{perfectly measured} (zero-uncertainty) events each.
    Each colour shows the results for an individual catalogue, while the true values are shown by gray dashed lines.
    The contours on the two-dimensional panels give the 90\% credible intervals.
    The legend gives the natural log one-dimensional Bayes factor and difference in maximum natural log likelihood comparing the gap and no-gap hypotheses for each mock catalogue, as well as the Bayes factor comparing the empty-gap and no-gap scenarios. Note that while the other four catalogues have a small amount of posterior support near $A=0$, the purple catalogue's support for $A=0$ is vanishingly small, explaining its anomalously high values of $\ln \mathcal{B}_A$ and $\ln \mathcal{B}_\mathrm{empty}$. This can be seen in the log-scale depth posterior in Appendix~\ref{sec:additional_results}.}
    \label{fig:mock_single_corner}
\end{figure}

As mentioned above, there are two sources of statistical uncertainty in these population analyses.
The first is the fluctuation in the number of observed or expected events with true parameters falling into the mass gap, driven by the binomial distribution.
The second is the measurement uncertainty in the parameters of individual events, which depends on the SNR of each detection.
To further investigate how these two sources of uncertainty drive our inconclusive results, we repeat our population analyses, this time assuming that every event in each mock catalogue is perfectly measured with no measurement uncertainty (i.e., each individual event posterior is represented by a delta function at the true injected value).
We do this several times for each catalogue, starting with 50 events and increasing the number of events by 50 each time (such that we analyse the five perfectly measured catalogues with 50, 100, 150, 200, and 250 events).
This gives an upper-bound on how well we might be able to infer such a feature given precisely measured events -- when the dominant factor driving the uncertainty is the number of events rather than the individual event uncertainties.
In other words, this analysis provides an estimate on how many high SNR detections might be required to infer a gap.
We find that the gap can somewhat reliably be inferred with moderate-to-high confidence with 250 perfectly measured events.
We plot the posterior distributions for the gap-related hyper-parameters and list the metrics comparing the gap hypothesis to the no-gap hypothesis in Fig.~\ref{fig:mock_single_corner}.
We show the posteriors for the perfectly measured catalogues with 50-200 events in Appendix~\ref{sec:additional_results}.

\section{Discussion}\label{sec:discussion}
We find that the gravitational-wave data are consistent with the presence of a gap in BBH component masses in the range $10 M_\odot$ to $15 M_\odot$ \citep[as predicted by][]{Schneider_2023}, and are also consistent with a dearth of component masses between $14 M_\odot$ and $22 M_\odot$ \citep[as predicted by][]{Disberg_2023}.
However, there is no significant statistical preference for any such feature.
Quantitatively, we find a natural log one-dimensional Bayes factor of only $\ln \mathcal{B}_A = 0.1$ or a difference in the maximum natural log likelihood of $\Delta \ln \mathcal{L}_{\max} = 1.0$ favouring the presence of a gap of any depth, and a natural log Bayes factor of $\ln \mathcal{B}_\mathrm{empty} = 0.0$ showing no preference for the case of a completely empty gap ($A=1$) over the case of no gap at all ($A=0$).

This result is perhaps unsurprising.
\cite{Farah_2023}, by analysing featureless mock catalogues with a data-driven mass model, show that weak evidence for under-abundances in the BBH mass spectrum (like the purported dip sometimes found near $\approx 10 M_\odot - 15 M_\odot$ in data-driven analyses) can often appear due to Poisson noise.
Quantitatively, a population drawn from a featureless power-law produces similar spuriously inferred features $\approx 20 \%$ of the time.

We find that this dip (if it does exist in nature) is still unlikely to be resolvable by the end of O4.
With 250 observed events, the level of confidence at which the gap can be inferred varies across our five mock catalogues from completely indeterminate ($\ln \mathcal{B}_A = 0.1$) to a modest preference for a gap of any kind ($\ln \mathcal{B}_A = 2.1$).
Shifting to a comparison between the completely empty-gap scenario and the no-gap scenario for these five mock catalogues, we find that values of $\ln \mathcal{B}_\mathrm{empty}$ range from 0.1 to 3.5.
Moreover, since catalogs with larger total values of the Bayes factor would be expected to have larger values of this quantity over a subset of the events, the low values of $\ln \mathcal{B}_A = 0.1$ and $\ln \mathcal{B}_\mathrm{empty}= 0.0$ across the GWTC-3 events further reduce the probability of a discriminating result by the end of O4.
A confident measurement of such a feature seemingly requires a large number (potentially several hundred) of BBH observations with very well measured masses.
Specifically, in the limiting case that individual-event mass posteriors have zero uncertainty, we find that a catalogue of 250 such events begins to reliably exhibit moderate-to-strong evidence for the presence of a gap, although even in this case, the Bayesian Information Criterion can fail to match the \cite{KassRaftery:1995} metric for detection once the penalty due to the extra parameters in the gap model is applied.
As such, we likely need on the order of $\sim 1000$ BBH detections (with uncertainty) to detect such a feature.
This is also in line with our estimate based off of the number of``missing'' in-gap events that would be required to overcome typical event-level uncertainties (see Section~\ref{sec:future}).
While a catalogue of this many events is beyond the purview of O4, such a catalogue may become available by the end of the fifth LVK observing run O5 \citep[see, for example][]{Petrov_2022, Kiendrebeogo_2023}.

We find that population inference can struggle to distinguish between a narrow gap from $\approx 10 M_\odot - 15 M_\odot$, and a wide, shallow dearth of masses between the two well measured peaks at $\approx 10 M_\odot$ and $\approx 30 M_\odot$ (see the covariance between $A$ and $\gamma_\mathrm{high}$ in Fig.~\ref{fig:corner} and Fig.~\ref{fig:mock_corner} for example).
This may indicate that we require a large number of BBH mergers with component masses between the gap's upper edge and the lower edge of the $\approx 30 M_\odot$ peak's tail (i.e. an abundance of observations around $20 M_\odot$).

Following this, if we adopt more confidence in the prior on the gap's location (restricting the gap edges to lie between $\gamma_\mathrm{low} = 10^{+1}_{-1}$ and $\gamma_\mathrm{low} = 15^{+1}_{-1}$), we find that the evidence for a gap in GWTC-3 increases.
Although still inconclusive, we find our natural log Bayes factor increases from $\ln \mathcal{B}_A = 0.1$ to $\ln \mathcal{B}_A = 0.6$, and from $\ln \mathcal{B}_\mathrm{empty} = 0.0$ to $\ln \mathcal{B}_\mathrm{empty} = 1.0$.
Similarly, in our mock catalogues, the preference for a gap tends to go from a typical value on the order of $\ln \mathcal{B}_A \sim 0.1$ up to a value on the order of $\ln \mathcal{B}_A \sim 1.0$ (with a similar increase in $\mathcal{B}_\mathrm{empty}$) when restricting the gap location -- though, this is still far from the \cite{KassRaftery:1995} Bayesian Information criterion threshold. 

In order to make a prediction for the chirp mass distribution using their model for the black hole component mass distribution, \cite{Schneider_2023} assume that black holes can only pair with other black holes on the same side of their hypothesised mass gap.
This follows from the understanding that merging black holes in isolated binaries should preferentially form with similar masses \citep[e.g.,][]{Mandel_2022}, although binary population synthesis models that do not account for a mass gap in black holes because of supernova physics predict a broader range of mass ratios \citep[see, e.g., Fig.~4 of][]{Broekgaarden_2022}.
Meanwhile, there may be observational evidence that, despite the distribution of mass ratios peaking at (or close to) equal masses ($q \approx 1$), a substantial fraction of the BBH population merges with unequal masses \citep[see, for example, data driven modelling of the mass ratio distribution in][but see also \citealt{Hoy_2024} for an alternative interpretation]{Callister_2023, Edelman_2023}. 
As we directly probe the component mass distribution for a gap, we are free to relax this assumption about mass pairings, instead inferring the degree to which BBH systems prefer equal masses from the data.
In reference to Fig.~\ref{fig:full_corner}, using GWTC-3 data, we find our inference on the mass-pairing power-law index $\beta$ (see equation~\ref{eq:mass_pairing}) is almost identical regardless of whether a gap is present ($\beta = 2.5^{+1.6}_{-1.1}$) or not ($\beta = 2.6^{+1.6}_{-1.1}$).
These measurements indicate a moderate preference for equal mass binaries along with broad support for unequal mass binaries, and is consistent with previous studies \citep[namely,][]{GWTC3_rnp, Farah_2023b}. 
On the other hand, it is conceivable that our chosen functional form for the mass and mass ratio distribution may represent a model misspecification which could contribute to a loss of discrimination ability for a mass gap.

There are a number of alternative mass-gap models that may be worth investigating in the future.
As simple extensions to the model used here, one may consider varying the structure of the mass distribution outside of the gap \citep[for example,][]{GWTC3_rnp}, the structure of the pairing function in equation~\ref{eq:mass_pairing} \citep[see, for example, the mass ratio distribution in][]{Edelman_2023}, and allowing for independent structure in the primary and secondary mass distributions \citep[different peaks and power-laws, for example;][]{Farah_2023b}.
Furthermore, one may be able to test for a gap by first fitting a carefully constructed data-driven model \citep[modelling component masses using the methods in, for example,][]{Edelman_2023, Callister_2023}, then testing for multi-modality in the region of the gap using a \cite{Hartigan_1985} dip test.
However, one must take care in using such a metric, as the BBH mass distribution has already been shown to be multi-modal in regions surrounding the proposed gap \citep{GWTC3_rnp, Edelman_2023, Callister_2023, Farah_2023}.

As a final note related to the above discussion, \cite{Galaudage_2024} also present an empirical analysis of the claims of \cite{Schneider_2023}, in a manuscript released shortly after this work entered review.
\cite{Galaudage_2024} similarly model the BBH primary mass distribution using a power-law structure, with two additional peaks.
In their work, however, the two peaks both have variable high and low-mass cutoffs which are fit to the data.
They assume that black hole masses arising from a \cite{Schneider_2023}-like formation scenario occupy the two peaks in their model, while the underlying power-law is considered to be pollution from other formation channels.
In doing so, they find that the two peaks prefer to sharply fall off in the region of the gap anticipated by \cite{Schneider_2023}, with the lower-mass peak cutting off at $\approx 11 M_\odot$, and the higher-mass peak turning on shortly after at $\approx 13 M_\odot$.
This leaves the $\approx 11 M_\odot - 13 M_\odot$ mass range occupied solely by ``pollution'' from the underlying power-law.
Regardless, \cite{Galaudage_2024} show qualitatively similar results for GWTC-3 as are presented in this work.
Namely, due to uncertainty in these peaks' inferred cutoffs \citep[see, for example, how the two peaks do indeed overlap within a portion of the 90\% credible intervals in Fig.~3 of][]{Galaudage_2024}, they do not find evidence for a notable under-abundance of black holes with masses in the range anticipated by \cite{Schneider_2023}.

\section*{Acknowledgements}
We thank Amanda Farah and our anonymous reviewer for their helpful comments on this manuscript.
We acknowledge support from the Australian Research Council (ARC) Centre of Excellence CE170100004 and ARC DP230103088.
This material is based upon work supported by NSF's LIGO Laboratory which is a major facility fully funded by the National Science Foundation.
The authors are grateful for computational resources provided by the LIGO Laboratory and supported by National Science Foundation Grants PHY-0757058 and PHY-0823459.

This research has made use of data or software obtained from the Gravitational Wave Open Science Center (gw-openscience.org), a service of LIGO Laboratory, the LIGO Scientific Collaboration, the Virgo Collaboration, and KAGRA. LIGO Laboratory and Advanced LIGO are funded by the United States National Science Foundation (NSF) as well as the Science and Technology Facilities Council (STFC) of the United Kingdom, the Max-Planck-Society (MPS), and the State of Niedersachsen/Germany for support of the construction of Advanced LIGO and construction and operation of the GEO600 detector. Additional support for Advanced LIGO was provided by the Australian Research Council. Virgo is funded, through the European Gravitational Observatory (EGO), by the French Centre National de Recherche Scientifique (CNRS), the Italian Istituto Nazionale di Fisica Nucleare (INFN) and the Dutch Nikhef, with contributions by institutions from Belgium, Germany, Greece, Hungary, Ireland, Japan, Monaco, Poland, Portugal, Spain. The construction and operation of KAGRA are funded by Ministry of Education, Culture, Sports, Science and Technology (MEXT), and Japan Society for the Promotion of Science (JSPS), National Research Foundation (NRF) and Ministry of Science and ICT (MSIT) in Korea, Academia Sinica (AS) and the Ministry of Science and Technology (MoST) in Taiwan.

\appendix

\section{Subtleties in modelling the chirp mass distribution}\label{sec:chirp_mass}
\begin{figure*}
    \centering
    \includegraphics[width=0.95\textwidth]{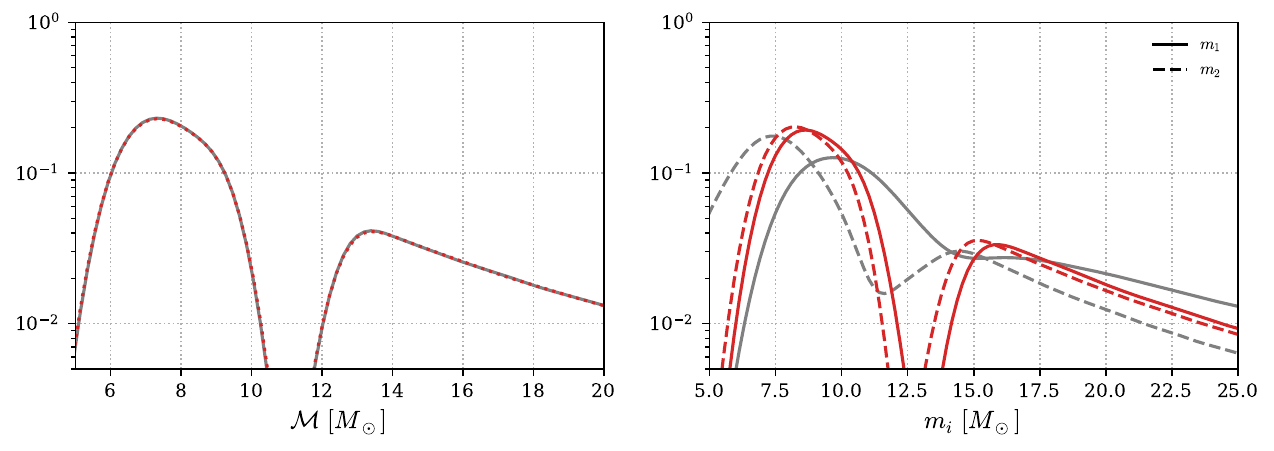}
    \caption{Toy gap models with (red) and without (gray) a strong preference for mass ratios near unity. The left panel shows identical power-law chirp mass distributions with gaps at $\approx 10 M_\odot - 12 M_\odot$ for both models.
    The right panel compares the distributions for primary (solid) and secondary (dashed) masses, corresponding to the two models.}
    \label{fig:mc_demo}
\end{figure*}

In this section, we demonstrate how a single chirp mass distribution can imply very different features in the component mass distributions under varying assumptions about the mass pairings.
We show this using two toy models in Fig.~\ref{fig:mc_demo}.
In both cases, we model the chirp mass distribution using a simple descending power-law with an added gap between $10 M_\odot$ and $12 M_\odot$ (see the notch filter described in Section~\ref{sec:model}).
In one model (gray), the distribution for mass ratios $q = m_2/m_1$ follows a power-law with a spectral index of $\beta=2$ \citep[a typical model and value inferred from the gravitational-wave data;][see also Fig.~\ref{fig:full_corner} below]{GWTC3_rnp}.
In the second model (red), we assume a much sharper spectral index of $20$, so that binary black holes can only be paired with very similar masses \citep[effectively disallowing black holes from pairing across a gap, as per][]{Schneider_2023}.
We then convert these distributions for $(\mathcal{M},q)$ to distributions for component masses $(m_1,m_2)$.
We see that the chirp mass gap produces a clear and substantial gap in component masses around the expected range of $10 M_\odot - 15 M_\odot$ when mass pairings are forced to be very similar.
However, when we allow for a broader spread in the mass ratios, we find a much shallower dip over a larger range in the component mass distribution.

\section{Mass distribution formalism}\label{sec:pi_m}
Here, we define the one-dimensional mass model used in this study:
\begin{align}\label{eq:1d}
    \nonumber
    \bar{\pi}(m_i|\Lambda) \propto &
    \Bigg[
    (1 - \lambda)\mathcal{P}(m_i|\alpha) +
    \lambda \Big(
    \lambda_1 \mathcal{N}_t(m_i|\mu_1,\sigma_1,m_{\min},m_{\max}) +
    (1 - \lambda_1) \mathcal{N}_t(m_i|\mu_2,\sigma_2,m_{\min},m_{\max})
    \Big)\Bigg] \times \\ &
    n(m_i|A,\gamma_\mathrm{low},\gamma_\mathrm{high},\eta_\mathrm{low},\eta_\mathrm{high})
    \Theta(m_{\max} - m_i) S(m_i|m_{\min}, \delta_m).
\end{align}
Here, $\Lambda$ denotes the set of all hyper-parameters that control the shape of the distribution, while $\mathcal{P}(m_i|\alpha)$ is a power-law with spectral index $\alpha$ and $\mathcal{N}_t(m_i|\mu,\sigma,m_\mathrm{low},m_\mathrm{high})$ denotes a truncated normal distribution with mean $\mu$, width $\sigma$, lower-bound $m_\mathrm{low}$, and upper-bound $m_\mathrm{high}$.
Meanwhile,
\begin{equation}
    n(m_i|A,\gamma_\mathrm{low},\gamma_\mathrm{high},\eta_\mathrm{low},\eta_\mathrm{high})
    = 1 - A \left[
    \left(1 + \left(\frac{m_i}{\gamma_\mathrm{low}}\right)^{\eta_\mathrm{low}}\right)
    \left(1 + \left(\frac{\gamma_\mathrm{high}}{m_i}\right)^{\eta_\mathrm{high}}\right)
    \right]^{-1}
\end{equation}
is a notch filter, which produces a gap of depth $A$, with a lower-bound $\gamma_\mathrm{low}$, upper-bound $\gamma_\mathrm{high}$, and parameters that control the sharpness of the roll-off on either side $\eta_\mathrm{low}$ and $\eta_\mathrm{high}$.
Finally, $\Theta(m_{\max} - m_i)$ is a Heaviside step-function which truncates the distribution at high masses, and
\begin{equation}
    S(m_i|m_{\min},\delta_m) = 
    \left[\exp\left(\frac{\delta_m}{m_i - m_{\min}} - \frac{\delta_m}{m_i-m_{\min}-\delta_m}\right) + 1\right]^{-1},
\end{equation}
truncates the distribution at low masses with a smoothing length $\delta_m$.

In Table~\ref{tab:priors}, we list all hyper-parameters that govern the mass distribution in this model, along with their associated priors adopted for hierarchical inference.

\begin{table*}
    \centering
    \begin{tabular}{|c l l l|}
        \hline
        \multicolumn{4}{|l|}{\textsc{Power-Law}}\\
        \hline
        $\alpha$ & $\mathcal{U}(-4, 12)$ & &
        Spectral-index of power-law component \\
        \hline
        \hline
        \multicolumn{4}{|l|}{\textsc{Peaks}}\\
        \hline
        $\lambda$ & $\mathcal{U}(0, 1)$ & &
        Fraction of masses in Gaussian peaks \\
        \hline
        $\lambda_1$ & $\mathcal{U}(0, 1)$ & &
        Fraction of peak-masses in lower-mass peak \\
        \hline
        $\mu_1$ & $\mathcal{U}(5, 20) $ & $[M_\odot]$ &
        Location of lower-mass peak \\
        \hline
        $\sigma_1$ & $\mathcal{U}(1, 5) $ & $[M_\odot]$ &
        Width of lower-mass peak \\
        \hline
        $\mu_2$ & $\mathcal{U}(20, 50) $ & $[M_\odot]$ &
        Location of upper-mass peak \\
        \hline
        $\sigma_2$ & $\mathcal{U}(1, 10) $ & $[M_\odot]$ &
        Width of upper-mass peak \\
        \hline
        \hline
        \multicolumn{4}{|l|}{\textsc{Gap}}\\
        \hline
        $\gamma_\mathrm{low}$ & $\mathcal{U}(\mu_1, \gamma_\mathrm{high}) $ & $[M_\odot]$ &
        Lower-edge location of gap \\
        \hline
        $\gamma_\mathrm{high}$ & $\mathcal{U}(\gamma_\mathrm{low}, \mu_2) $ & $[M_\odot]$ &
        Upper-edge location of gap \\
        \hline
        $\eta_\mathrm{low}$ & $\mathcal{U}(0, 50)$ & &
        Sharpness of gap's lower-edge \\
        \hline
        $\eta_\mathrm{high}$ & $\mathcal{U}(0, 50)$ & &
        Sharpness of gap's upper-edge \\
        \hline
        $A$ & $\mathcal{U}(0, 1)$ & &
        Depth of gap \\
        \hline
        \hline
        \multicolumn{4}{|l|}{\textsc{Cutoffs}}\\
        \hline
        $m_{\max}$ & $\mathcal{U}(60, 100) $ & $[M_\odot]$ &
        Maximum allowed mass \\
        \hline
        $m_{\min}$ & $\mathcal{U}(2, 10) $ & $[M_\odot]$ &
        Minimum allowed mass \\
        \hline
        $\delta_m$ & $\mathcal{U}(0, 10) $ & $[M_\odot]$ &
        Length of minimum-mass roll-off \\
        \hline
        \hline
        \multicolumn{4}{|l|}{\textsc{Pairing}}\\
        \hline
        $\beta$ & $\mathcal{U}(-4, 12)$ & &
        Power-law index of pairing function \\
        \hline
    \end{tabular}
    \caption{List of hyper-parameters used in the mass model, along with their priors and descriptions, grouped by feature.
    Here, $\mathcal{U}(a,b)$ indicates a uniform prior from $a$ to $b$.
    Of course, we enforce that the upper edge of the gap is higher than the lower edge of the gap $\gamma_\mathrm{high} > \gamma_\mathrm{low}$.
    We also enforce that the gap lies between the two Gaussian peaks, which are relatively precisely measured to lie near $\approx 10 M_\odot$ and $\approx 30 M_\odot$ \citep{GWTC3_rnp}.
    We do so to avoid model aberrations from overlapping gaps and peaks.
    }
    \label{tab:priors}
\end{table*}

\section{Additional results} \label{sec:additional_results}

\begin{figure*}
    \centering
    \includegraphics[width=0.7\columnwidth]{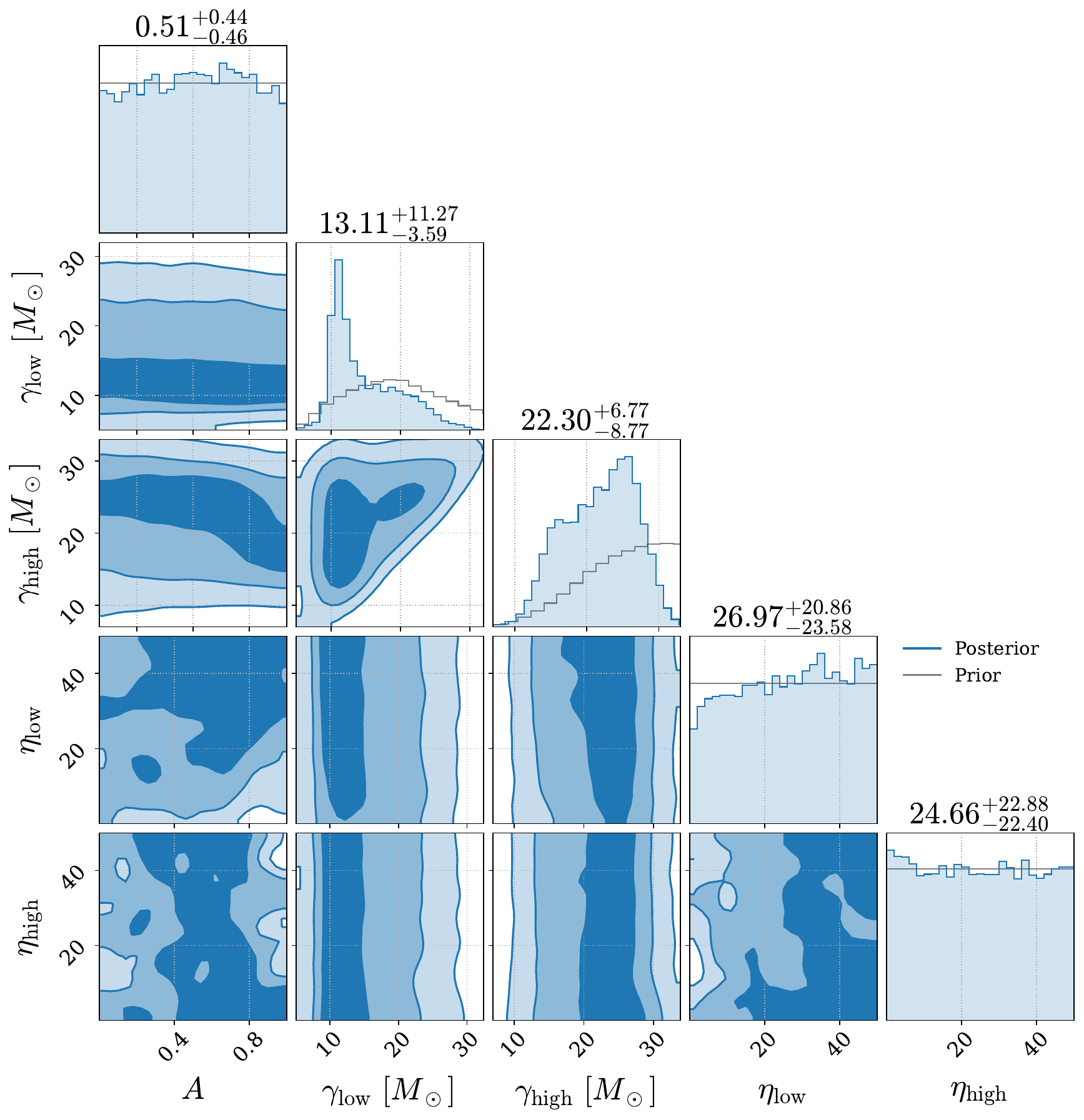}
    \caption{Posterior distributions for gap-related population hyper-parameters inferred using GWTC-3 data, including the edge sharpness.
    The values listed above give the median and 90\% credible intervals on these posteriors.
    The contours on the two-dimensional panels give the 50\%, 90\%, and 99\% credible intervals.
    The priors are over-plotted in gray.}
    \label{fig:corner_eta}
\end{figure*}

\begin{figure*}
    \centering
    \includegraphics[width=\columnwidth]{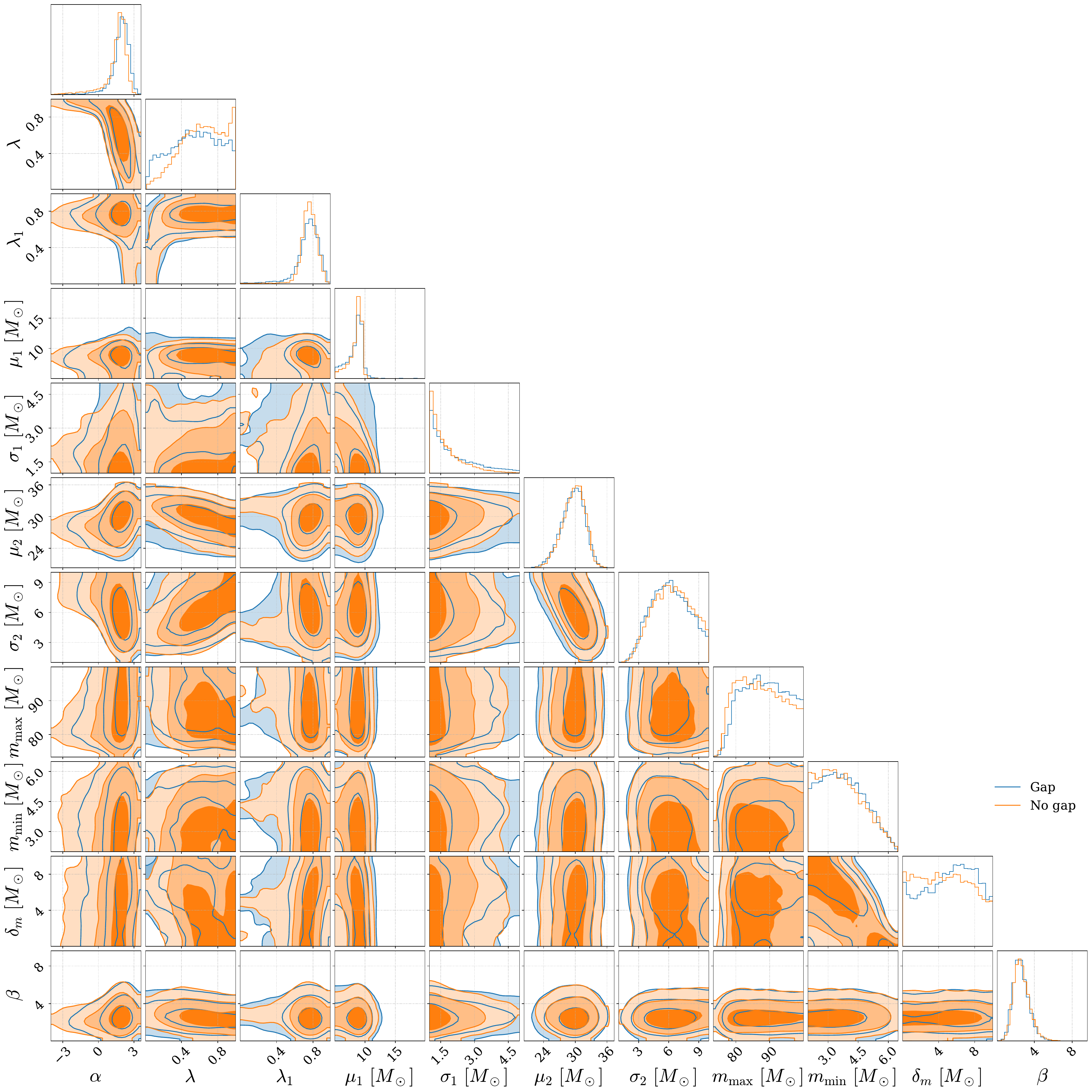}
    \caption{Posterior distributions for mass-related population hyper-parameters inferred using GWTC-3 data.
    Blue shows the results when a gap is included, while orange shows the results when a gap is not included.
    The contours on the two-dimensional panels give the 50\%, 90\%, and 99\% credible intervals.
    }
    \label{fig:full_corner}
\end{figure*}

\begin{figure*}
    \centering
    \includegraphics[width=0.45\columnwidth]{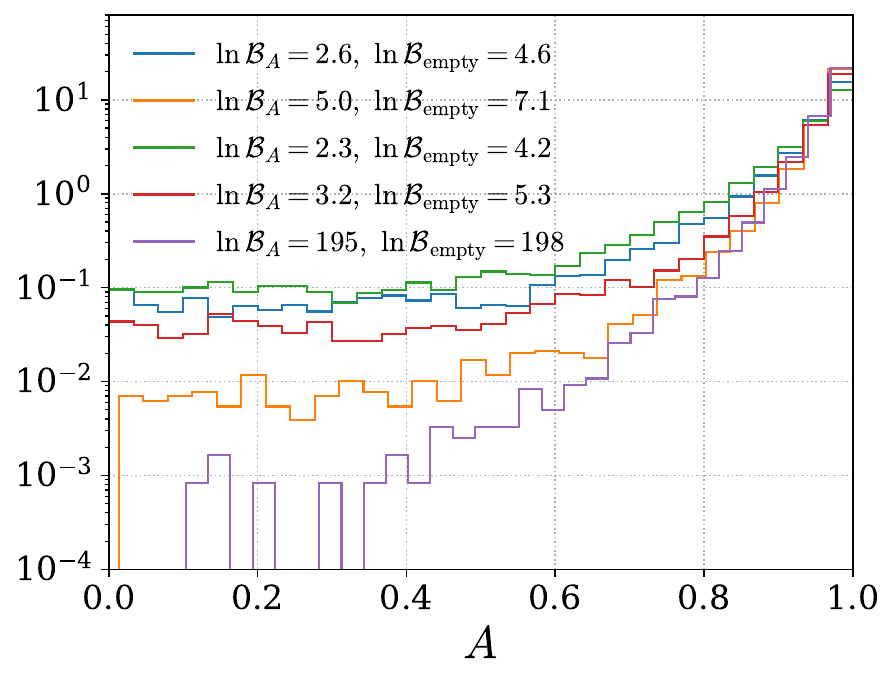}
    \caption{
    Log-scale posterior distributions for the gap depth inferred from five mock catalogues containing a mass gap, each with 250 perfectly measured events (see Fig.~\ref{fig:mock_single_corner}).
    Note that the posterior support for $A=0$ is vanishingly small in the purple catalogue, explaining the extreme values of $\ln \mathcal{B}_A$ and $\ln \mathcal{B}_\mathrm{empty}$.
    }
    \label{fig:mock_depth_single_250_log}
\end{figure*}

\begin{figure*}
    \centering
    \includegraphics[width=0.49\columnwidth]{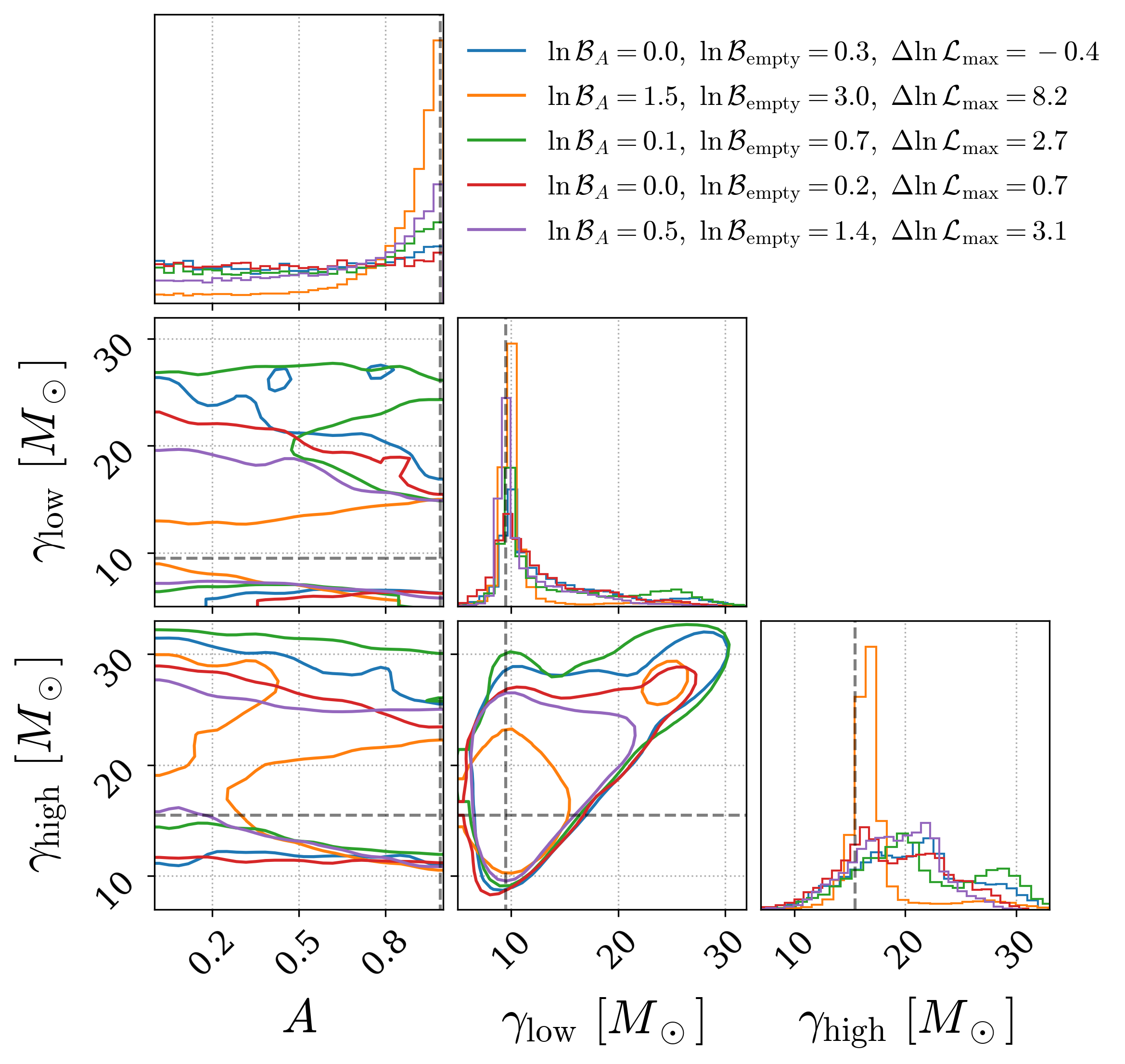}\quad\includegraphics[width=0.49\columnwidth]{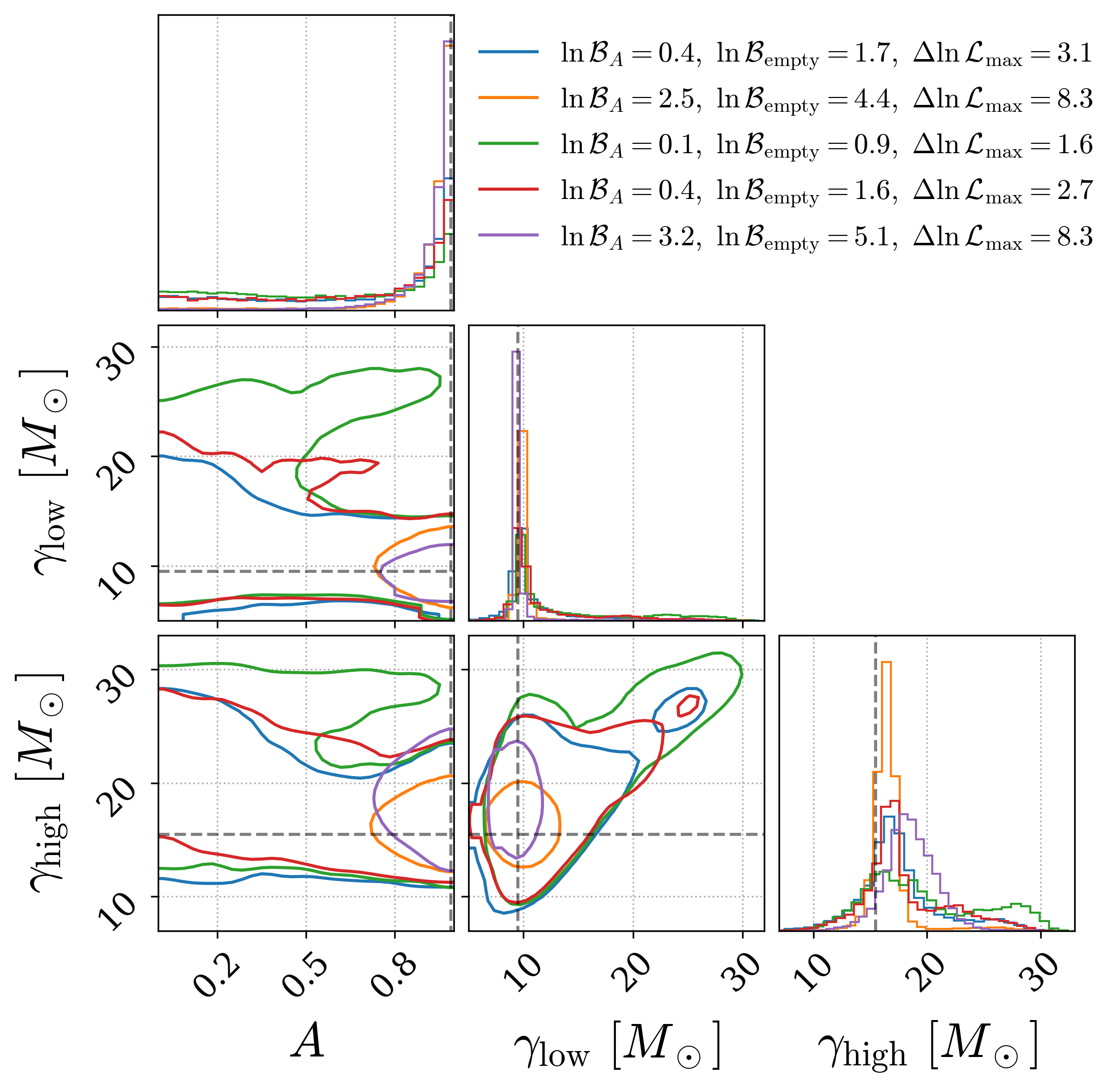}
    \\ \vspace{10mm}
    \includegraphics[width=0.49\columnwidth]{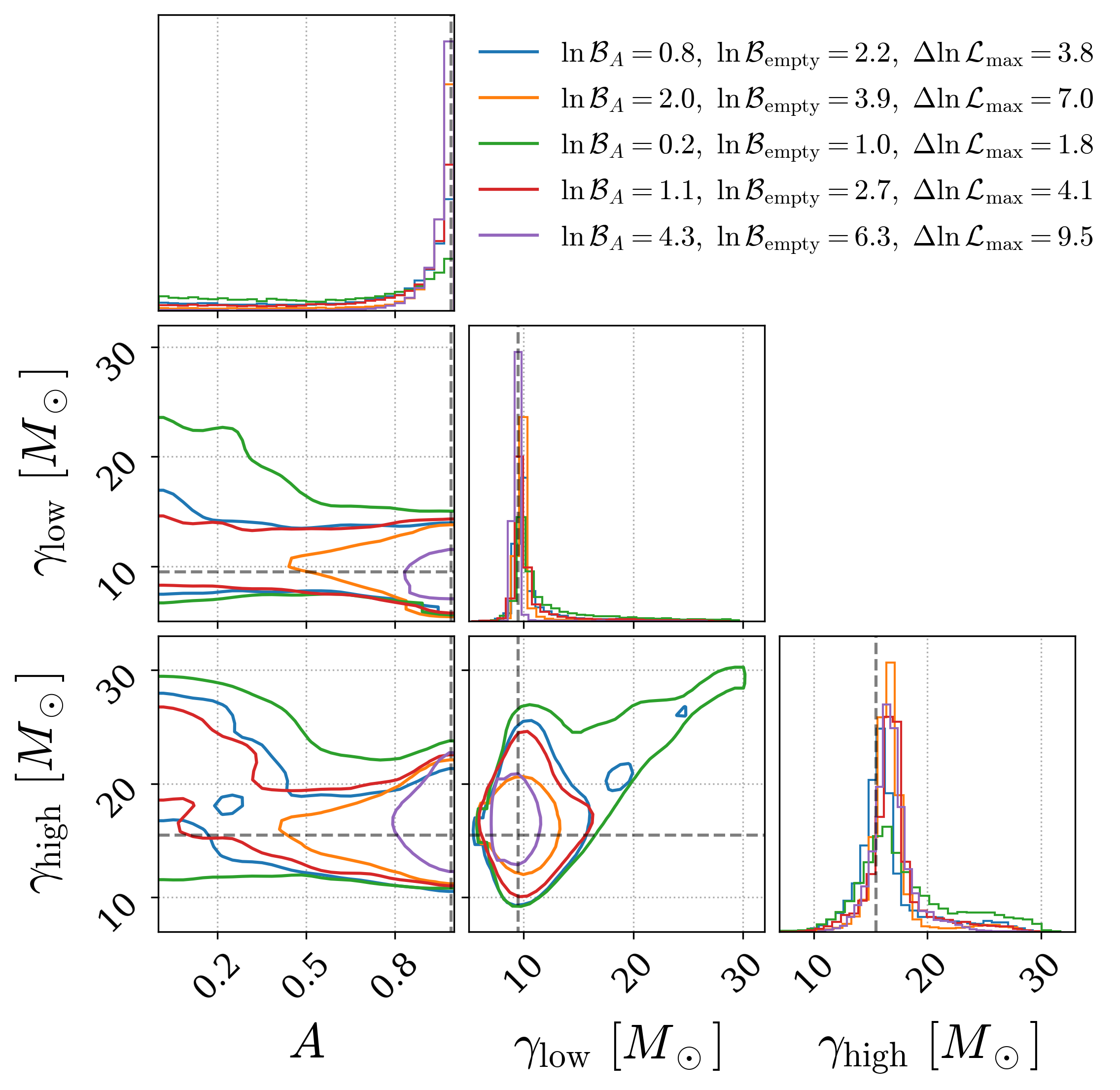}\quad\includegraphics[width=0.49\columnwidth]{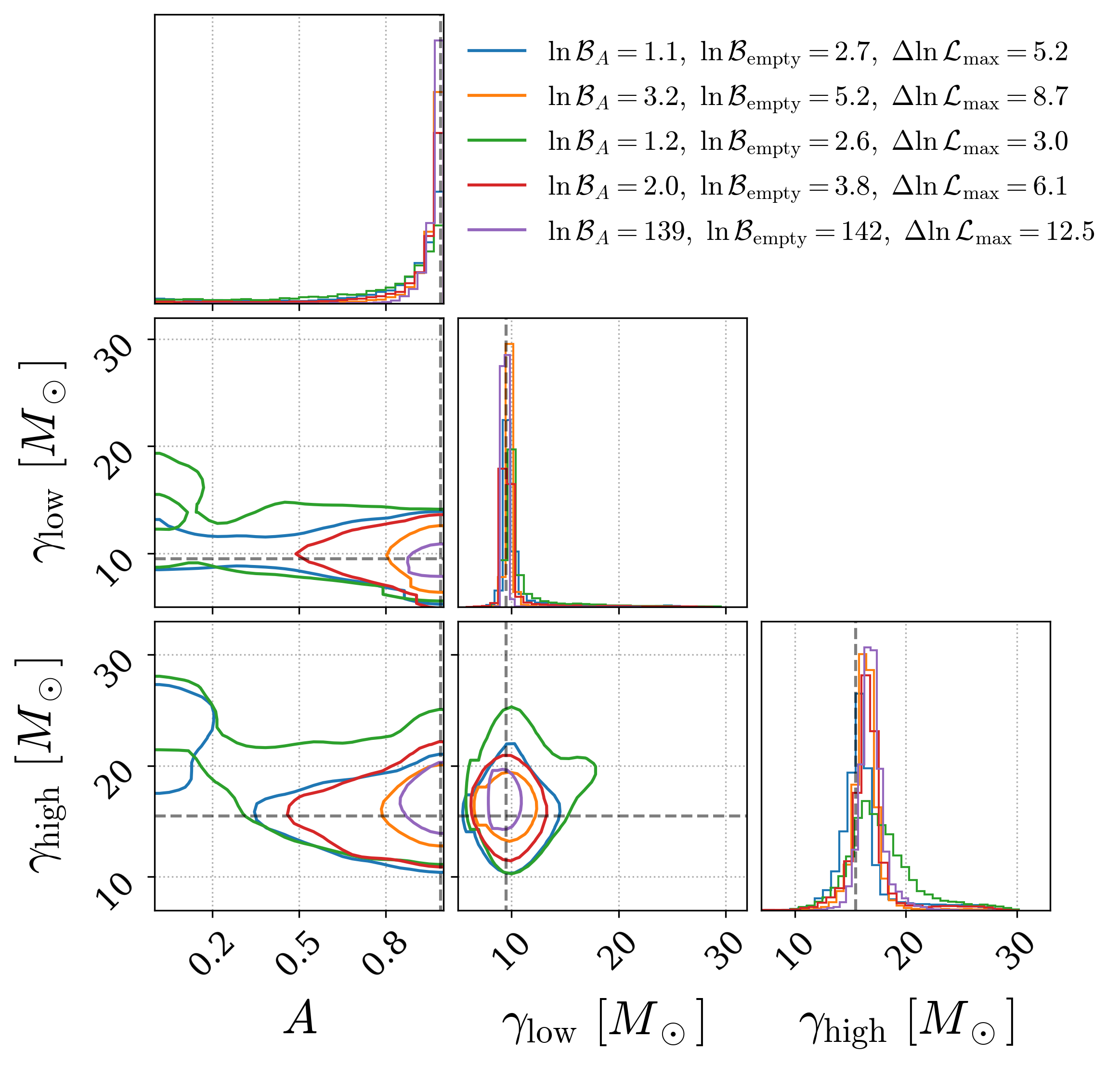}
    \caption{Posterior distributions for gap-related population hyper-parameters inferred from mock catalogues with 50, 100, 150, and 200 \textit{perfectly measured} (zero-uncertainty) events.
    Each colour shows the results for an individual catalogue, while the true values are shown by gray dashed lines.
    The contours on the two-dimensional panels give the 90\% credible intervals.
    The legends give the natural log one-dimensional Bayes factors and differences in maximum natural log likelihoods comparing the gap and no-gap hypotheses for each mock catalogue, as well as the natural log Bayes factor comparing the empty-gap to no-gap scenarios.
    Note that the vertical axes in the marginal posterior distributions are scaled differently for each corner plot.}
    \label{fig:mock_single_corner_var}
\end{figure*}

\begin{figure*}
    \centering
    \includegraphics[width=0.67\textwidth]{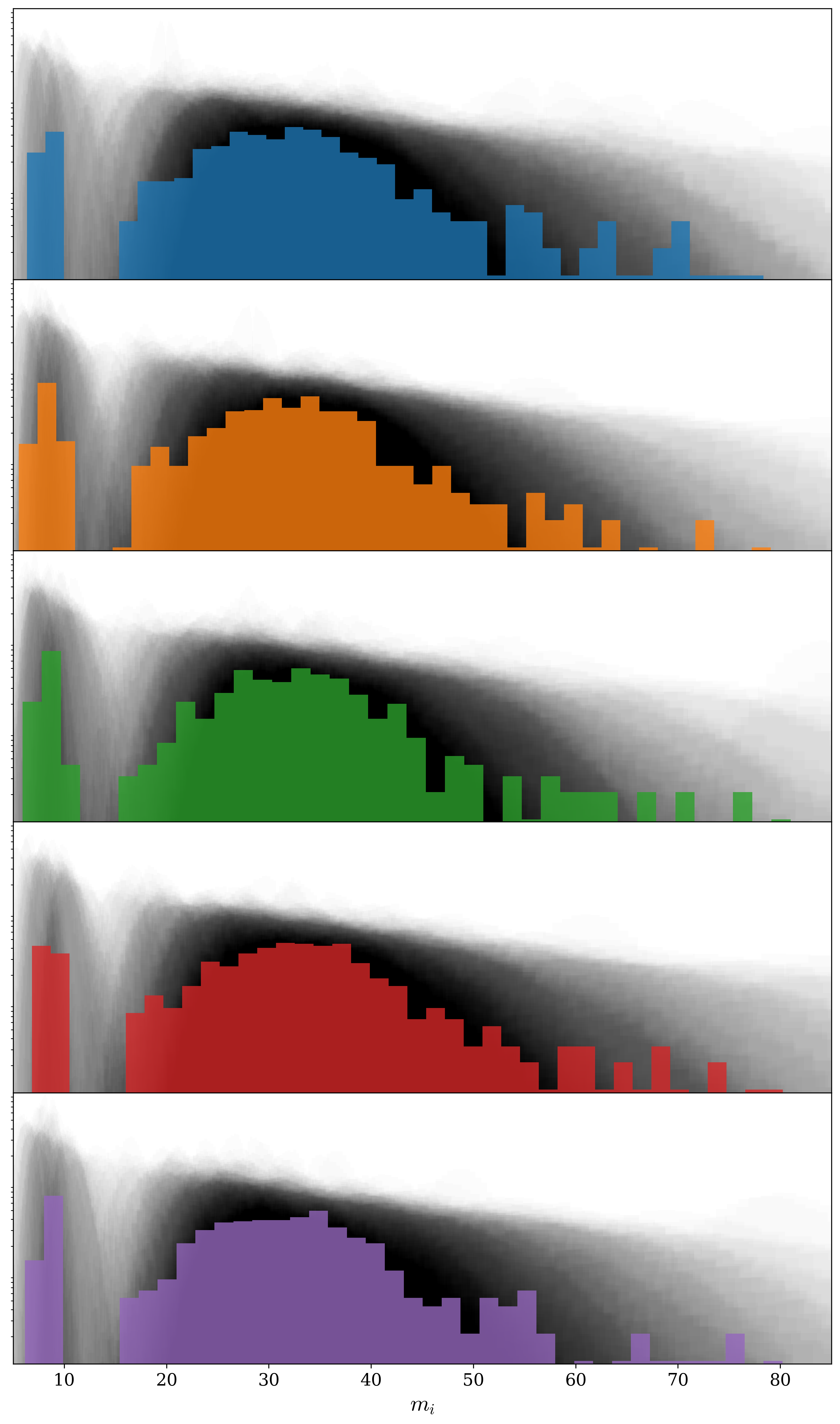}
    \caption{
    Log-scale histograms of the masses of all black holes (both $m_1$ and $m_2$) in each mock catalogue used in Section~\ref{sec:future}.
    The colored histograms show the true mass values for the events in a given catalogue, while the simulated mass posteriors associated with the events are shown in gray.
    Note that the cumulative density of the posterior distributions for all events in a given catalogue is not represented by the height of the gray distributions (these are normalized per-event).
    }
    \label{fig:cats_plot}
\end{figure*}

In this section, we show additional results that may be of interest to some readers.
In Fig.~\ref{fig:corner_eta}, we show the gap-related hyper-parameters, including the edge sharpness parameters $\eta_\mathrm{low}$ and $\eta_\mathrm{high}$, inferred from GWTC-3 data.
In Fig.~\ref{fig:full_corner}, we show the posteriors for all other mass-related hyper-parameters inferred from GWTC-3 data.
Fig.~\ref{fig:mock_depth_single_250_log} shows, in log scale, the inferred posterior depth for the five mock catalogues with 250 perfectly measured events (i.e., an expansion of the panel showing the marginal posterior for $A$ in Fig.~\ref{fig:mock_single_corner} but on a log scale).
The posterior distributions for the gap-related hyper-parameters for each mock catalogue with perfectly measured (zero-uncertainty) events, when using 50, 100, 150, and 200 events, are displayed in Fig.~\ref{fig:mock_single_corner_var}.
Finally, the masses for each event in each mock catalogue are plotted in Fig.~\ref{fig:cats_plot}.

\bibliographystyle{aasjournal}
\bibliography{refs}

\end{document}